\documentclass{appolb}
\usepackage{graphicx}
\usepackage[utf8]{inputenc}

\begin{document}
\title{$T$-dependence of the axion mass when the $U_A(1)$
 and chiral symmetry breaking are tied   %
\thanks{Presented at Excited QCD 2019, Schladming, Austria,
 Jan. 30 -- Feb. 3, 2019.}%
}
\author{Dubravko Klabu\v{c}ar, Davor Horvati\'{c}
\address{Physics Department, Faculty of Science -- PMF, University of Zagreb, Croatia}
\\
Dalibor Kekez
\address{Rudjer Bo\v{s}kovi\'c Institute, Zagreb, Croatia}
}
\maketitle
\begin{abstract}
\noindent   
 Modulo the scale of spontaneous breaking of Peccei-Quinn symmetry, the
 axion mass $m_{\rm a}(T)$ is given by the QCD topological susceptibility
 $\chi(T)$ at all temperatures $T$. From an approach tying the $U_A(1)$ and
 chiral symmetry breaking and getting good $T$-dependence of $\eta$ and 
 $\eta'$ mesons, we get $\chi(T)$ for an effective Dyson-Schwinger
 model of nonperturbative QCD. Comparison with lattice results for $\chi(T)$,
 and thus also for $m_{\rm a}(T)$, shows good agreement for temperatures ranging
 from zero up to the double of the chiral restoration temperature $T_c$.
\end{abstract}
  
\section{Introduction}

The fundamental theory of strong interactions, QCD, has the so-called Strong CP
problem. Namely, there is no experimental evidence of any CP-symmetry violation
in strong interactions, although there is in principle no reason why the QCD
Lagrangian should not include  the so-called $\Theta$-term
${\cal L}^{\Theta}$, where gluon fields $F^b_{\mu\nu}(x)$ comprise the
 CP-violating combination $Q(x)$,
\begin{equation}
{\cal L}^{\Theta}(x) \, = \, \Theta \, \frac{g^2}{\, 64\, \pi^2\, } \,
      \epsilon^{\mu\nu\rho\sigma} \, F^b_{\mu\nu}(x) \, F^b_{\rho\sigma}(x) \,
 \equiv \, \Theta \, Q(x) \, .
\label{Theta-term}
\end{equation}
 Admittedly, ${\cal L}^{\Theta}$ can be rewritten
as a total divergence, but, unlike in QED, this does not enable
 discarding it in spite of the gluon fields vanishing sufficiently fast as
 $|x| \to \infty$.  
This is because of nontrivial topological structures in QCD, such as instantons,
which are important for, {\it e.g.}, solving of the $U_A(1)$ problem
and yielding the anomalously large mass of the $\eta'$ meson.

Thus, there is no reason why the coefficient $\Theta$ of this term should be
of a very different magnitude from the coefficients of the other, CP-symmetric
 terms comprising the usual CP-symmetric QCD Lagrangian. Nevertheless, the
 experimental bound on the coefficient of the term is extremely low,
 $|\Theta| < 10^{-10}$ \cite{Baker:2006ts}, and consistent with zero.
 This is the mystery of the missing strong CP violation:
  why is  $\Theta$  so small?

Various proposed theoretical solutions stood the test of time with varying success.
A long-surviving solution, which is actually the preferred solution nowadays,
 is a new particle beyond the Standard Model - the axion. Important is also that
 axions turned out to be very interesting also for cosmology, as promising
candidates for dark matter. (See, {\it e.g.,} \cite{Wantz:2009it,Kim:2017tdk}.)

\section{Axion mass from the non-Abelian axial anomaly}

Peccei and Quinn introduced \cite{Peccei:1977hh,Peccei:1977ur} a new axial global
symmetry $U(1)_{PQ}$ which is broken spontaneously at some scale $f_{\rm a}$.
This presumably huge \cite{Tanabashi:2018oca} but otherwise presently unknown
scale is the key free parameter of axion theories, which determines the absolute
value of the axion mass $m_{\rm a}$. However, this constant cancels from
 combinations such as $m_{\rm a}(T)/m_{\rm a}(0)$. 
Hence, useful insights and applications are possible in spite of $f_{\rm a}$ being not known.

We have often, including applications at $T>0$  
\cite{Horvatic:2007qs,Horvatic:2007wu,Horvatic:2007mi,Benic:2011fv,Horvatic:2018ztu},
employed a chirally well-behaved relativistic bound-state approach to modeling
nonperturbative QCD through Dyson-Schwinger equations (DSE) for Green's functions
of the theory. (For reviews, see \cite{Alkofer:2000wg,Roberts:2000aa,Fischer:2006ub}
for example.) 
Such calculations can yield model predictions on the QCD topological susceptibility,
including its temperature dependence $\chi(T)$, which are correctly related to the
QCD dynamical chiral symmetry breaking (DChSB) and restoration. It turns out that
$\chi(T)$ is precisely that factor in the axion mass $m_{\rm a}(T)$, which carries
the nontrivial $T$-dependence.


\subsection{Axions as quasi-Goldstone bosons}

The pseudoscalar axion field  $\mbox{\large a}(x)$ arises as the
(would-be massless) Goldstone boson of the spontaneous breaking of the
 Peccei-Quinn symmetry \cite{Weinberg:1977ma,Wilczek:1977pj}.
The axion contributes to the total Lagrangian its kinetic term and its
interaction with the standard-model fermions. But what is important for
the resolution of the strong CP problem, is that the axion also couples
to the topological charge density operator operator $Q(x)$ in 
Eq. (\ref{Theta-term}). Then, the $\Theta$-term in the QCD Lagrangian
 gets modified to
\begin{equation}
{\cal L}^{\Theta \, +}_{\rm axion}  \, = \, {\cal L}^{\Theta} \, + 
 \frac{\mbox{\large a}(x)}{f_{\rm a}} \, Q(x)\,  
 = \, \left(\, \Theta \, + \frac{\mbox{\large a}}{f_{\rm a}} \right) \,
 \frac{g^2}{64\pi^2} \,
           \epsilon^{\mu\nu\rho\sigma} F^b_{\mu\nu}\, F^b_{\rho\sigma} \, .
\label{Theta+axion}
\end{equation}
Through this coupling of the axion to gluons, the $U(1)_{PQ}$ symmetry is also broken
{\it explicitly} by the $U_A(1)$ non-Abelian, gluon axial anomaly, so that the axion 
has a non-vanishing mass, $m_{\rm a} \neq 0$ \cite{Weinberg:1977ma,Wilczek:1977pj}.

Gluons generate an effective axion potential, and its minimization leads to 
the axion expectation value $\langle \mbox{\large a} \rangle$ such that the
modified coefficient, multiplying the topological charge density $Q(x)$,
 should vanish:
\begin{equation}
\Theta \, + \frac{\langle \mbox{\large a} \rangle}{f_{\rm a}} \, =\, 0 \, .
\label{minimizeEffV}
\end{equation}
 The strong CP problem is thereby solved, irrespective of the initial
 value of $\Theta$. Relaxation from any $\Theta$-value in
 the early Universe towards the minimum at Eq. (\ref{minimizeEffV}),
is known as misalignment production, and the the resulting axion oscillation energy
is a cold dark matter candidate ({\it e.g.}, see \cite{Wantz:2009it,Kim:2017tdk}).


\subsection{Axion mass from anomalous $U_A(1)$ breaking driven by DChSB}

A direct measure of the $U_A(1)$ symmetry breaking is the topological susceptibility
 $\chi$, given by the convolution of the time-ordered product $\cal T$ of the
 topological charge densities $Q(x)$ defined by Eq. (\ref{Theta-term}) [or
 Eq. (\ref{Theta+axion})]:
\begin{equation}
\chi \,  =  \,
\int d^4x \; \langle 0|\, {\cal T} \, Q(x) \, Q(0) \, |0 \rangle \; .
\label{chi}
\end{equation}
 The expansion of the effective axion 
potential reveals in its quadratic term that the axion mass squared
(times $f_{\rm a}^2$) is equal{\footnote{To a high level of accuracy, since 
corrections to Eq. (\ref{axionMass}) are of the order $M_\pi^2/f_{\rm a}^2$
\cite{Gorghetto:2018ocs}, where the pion mass $M_\pi$ is negligible.  }}
 to the QCD topological susceptibility. 
This holds for all temperatures $T$:
\begin{equation}
  m_{\rm a}^2(T) \, f_{\rm a}^2 \, =  \,  \chi(T) \, . 
\label{axionMass}
\end{equation}
On the other hand, in our study \cite{Horvatic:2018ztu} of the $T$-dependence
of the $\eta$ and $\eta'$ masses and the influence of the anomalous $U_A(1)$
 breaking and restoration, we used the light-quark-sector result
 \cite{Di Vecchia:1980ve,Leutwyler:1992yt,Durr:2001ty}
\begin{equation}
\chi(T) \, = \, \frac{ - \, 1}{ \,\,\, 
             \frac{1}{\, m_{u} \, \langle {\bar u}u \, (T) \rangle } +
          \frac{1}{\, m_{d}\,\langle{\bar d}d  \, (T) \rangle} +
        \frac{1}{\, m_{s}\,\langle {\bar s}s  \, (T) \rangle }
            \,\,\,  } \, + \, {\cal C}_m  \, 
\label{chiLIGHTq}
\end{equation}
where ${\cal C}_m$ is a very small correction term of higher orders in the
small current quark masses $m_q$ ($q=u,d,s$), and in the present context
we do not consider it further. Thus, the overwhelming part, namely the
leading term of $\chi$, is given by the quark condensates
$\langle {\bar q}q \rangle$ ($q=u,d,s$), which arise as order parameters
 of DChSB. Their temperature dependence determines that of $\chi(T)$,
which in turn determines the $T$-dependence of the anomalous part of
the pseudoscalar meson masses in the $\eta$-$\eta'$ complex.
This is the mechanism of Ref. \cite{Horvatic:2018ztu}, how DChSB and
chiral restoration drive, respectively, the anomalous breaking and
restoration of the $U_A(1)$ symmetry of QCD.

Now, Eqs. (\ref{axionMass}) and (\ref{chiLIGHTq}) show that
this mechanism determines also the $T$-dependence of the axion mass.

\begin{figure}[htb]
\centerline{%
\includegraphics[width=12.5cm]{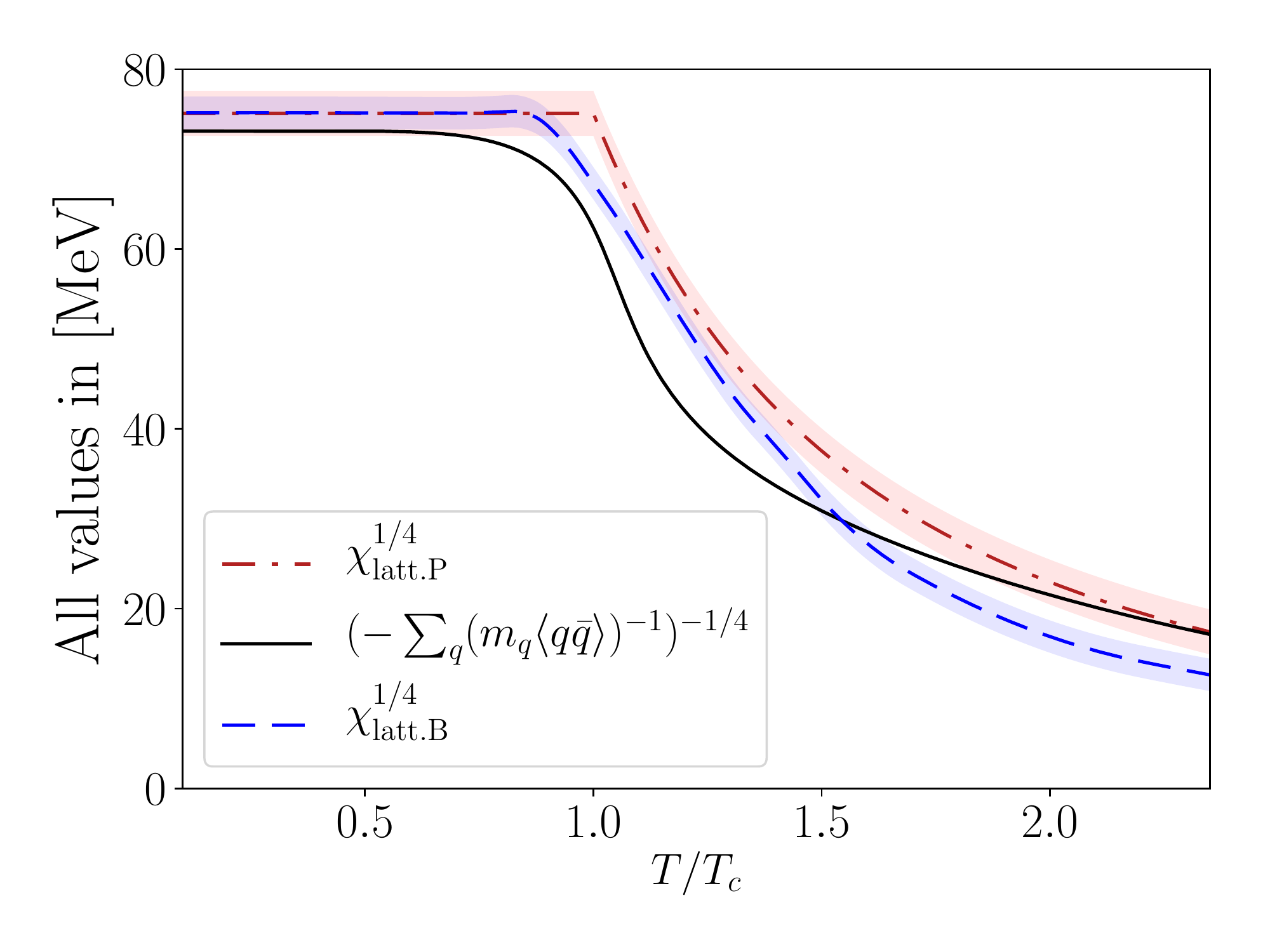}}
\caption{The relative temperature $T/T_c$ dependence of (the leading term of)
 $\chi(T)$ from our oft-adopted
\cite{Horvatic:2007qs,Horvatic:2007wu,Horvatic:2007mi,Benic:2011fv,Horvatic:2018ztu}
chirally well-behaved DSE model (solid curve), and from lattice:
dash-dotted curve extracted from Petreczky {\it et al.,} \cite{Petreczky:2016vrs}
and dashed curve extracted from Borsany {\it et al.,} \cite{Borsanyi:2016ksw}).
(Colors online.)}
\label{Fig:F2H}
\end{figure}

To describe $\eta'$ and $\eta$ mesons, it is essential to include $U_A(1)$
symmetry breaking at least at the level of the masses. This could be done
simply \cite{Klabucar:1997zi,Kekez:2005ie,Benic:2014mha}, by adding the
anomalous contribution to isoscalar meson masses as a perturbation, thanks
to the fact that the $U_A(1)$ anomaly is suppressed in the limit of large
number of QCD colors $N_c$ \cite{Witten:1979vv,Veneziano:1979ec}. Concretely,
Ref. \cite{Benic:2014mha} adopted Shore's equations \cite{Shore:2006mm},
where the $U_A(1)$-anomalous contribution to the light pseudoscalar masses
is expressed through the condensates of light quarks with non-vanishing
current masses. They are thus used also in $\chi(T)$ (\ref{chiLIGHTq}), since
this approach has recently been extended \cite{Horvatic:2018ztu} to $T>0$.
This gave us our results for $\chi(T)$ depicted in Fig. \ref{Fig:F2H}.

Indeed, the now established smooth, crossover behavior around
the pseudocritical temperature $T_c$ for the chiral transition,
is obtained for the DChSB condensates of realistically massive
light quarks -- {\it i.e.,} the quarks with realistic
explicit chiral symmetry breaking \cite{Horvatic:2018ztu}.
In contrast, using in Eq. (\ref{chiLIGHTq}) the massless quark condensate
$\langle {\bar q}q \rangle_0$ (which drops sharply to zero at $T_c$)
instead of the ``massive'' ones,
would dictate a sharp transition of the second order at $T_c$
\cite{Benic:2011fv,Horvatic:2018ztu} also for $\chi(T)$.
Obviously, this would imply that axions are massless for $T > T_c$.

In Fig. \ref{Fig:F2H}, we present (the leading term of) our model-calculated
 \cite{Horvatic:2018ztu} $\chi(T)^{1/4}$, depicted as the solid curve. Due 
to Eq. (\ref{axionMass}), this is our model prediction for
 $\sqrt{m_{\rm a}(T) \, f_{\rm a}}$. For temperatures
up to $T\approx 2.3\, T_c$, we compare it to the lattice results of Petreczky
{\it et al.,} \cite{Petreczky:2016vrs} and of Borsany {\it et al.,}
 \cite{Borsanyi:2016ksw}, rescaled to the relative temperature $T/T_c$.


\section{Summary}

 The axion mass and its temperature dependence $m_{\rm a}(T)$
 can be calculated in an effective model of nonperturbative QCD
 (up to the constant scale parameter $f_{\rm a}$) as the square root
 of the topological susceptibility $\chi(T)$. We obtained
 it from the condensates of $u$-, $d$- and $s$-quarks and antiquarks
 calculated in the SDE approach using a simplified nonperturbative 
 model interaction \cite{Horvatic:2018ztu}. 
 Our prediction on $m_{\rm a}(T)$ is thus supported by the fact
 that our topological susceptibility also yields the $T$-dependence
 of the $U_A(1)$ anomaly-influenced masses of $\eta'$ and $\eta$
 mesons which is consistent with experimental evidence \cite{Horvatic:2018ztu}.

 Our result on $\chi(T)$ and the related axion mass is qualitatively
 similar to the one obtained in the framework of the NJL model \cite{Lu:2018ukl}. 
Our topological susceptibility is also qualitatively similar to the
 pertinent lattice results \cite{Petreczky:2016vrs,Borsanyi:2016ksw},
 except that our dynamical model could so far access only much smaller
 range of temperatures, $T < 2.4\, T_c$. 
 On the other hand, the lattice supports the smooth crossover transition
 of $\chi(T)$, which is, in our approach, the natural consequence of
 employing the massive-quark condensates exhibiting crossover around the
 chiral restoration temperature $T_c$. Hence, the (partial) $U_A(1)$
 restoration observed in Ref. \cite{Horvatic:2018ztu} must also be a
 crossover, which in the present work, as well as in its longer
 counterpart \cite{Horvatic:2019eok} containing a detailed analysis
 of the model parameter dependence, translates into the corresponding
 smooth $T$-dependence of the axion mass.

\end{document}